\documentclass[acmtog,screen,nonacm]{acmart}

\acmSubmissionID{770}

\usepackage{booktabs} 

\usepackage[ruled]{algorithm2e} 
\usepackage{multirow}
\usepackage{overpic}
\newcommand{\revision}[1]{\textcolor{black}{#1}}

\SetAlFnt{\small}
\SetAlCapFnt{\small}
\SetAlCapNameFnt{\small}
\SetAlCapHSkip{0pt}

\acmJournal{TOG}

\begin{document}
\title{
SAND: Spatially Adaptive Network Depth for Fast Sampling of Neural Implicit Surfaces}

\author{Chuanxiang Yang}
\affiliation{\institution{Shandong University} 
\country{China}}
\email{chxyang2023@gmail.com}

\author{Junhui Hou}
\affiliation{\institution{City University of Hong Kong} 
\country{China}}
\email{jh.hou@cityu.edu.hk}

\author{Yuan Liu}
\affiliation{\institution{Hong Kong University of Science and Technology} 
\country{China}}
\email{yuanly@ust.hk}

\author{Siyu Ren}
\affiliation{\institution{City University of Hong Kong} 
\country{China}}
\email{siyuren2-c@my.cityu.edu.hk}

\author{Guangshun Wei}
\affiliation{  \institution{Shandong University}
\country{China}}
\email{guangshunwei@gmail.com}

\author{Taku Komura} 
\affiliation{  \institution{The University of Hong Kong}
\country{China}}
\email{taku@cs.hku.hk}

\author{Yuanfeng Zhou}
\authornote{Corresponding authors: Yuanfeng Zhou and Wenping Wang.}
\affiliation{\institution{Shandong University} 
\country{China}}
\email{yfzhou@sdu.edu.cn}

\author{Wenping Wang}
\authornotemark[1]
\affiliation{  \institution{Texas A\&M University}
\country{USA}}
\email{wenping@tamu.edu}

\begin{abstract}
Implicit neural representations are powerful for geometric modeling, but their practical use is often limited by the high computational cost of  network evaluations. We observe that implicit representations require progressively lower accuracy as query points move farther from the target surface, and that even within the same iso-surface, representation difficulty varies spatially with local geometric complexity. However, conventional neural implicit models evaluate all query points with the same network depth and computational cost, ignoring this spatial variation and thereby incurring substantial computational waste. Motivated by this observation, we propose an efficient neural implicit geometry representation framework with spatially adaptive network depth (SAND). SAND leverages a volumetric network-depth map together with a tailed multi-layer perceptron (T-MLP) to model implicit representation. The volumetric depth map records, for each spatial region, the network depth required to achieve sufficient accuracy, while the T-MLP is a modified MLP designed to learn implicit functions such as signed distance functions, where an output branch, referred to as a tail, is attached to each hidden layer. This design allows network evaluation to terminate adaptively without traversing the full network and directs computational resources to geometrically important and complex regions, improving efficiency while preserving high-fidelity representations. Extensive experimental results demonstrate that our approach can significantly improve the inference-time query speed of implicit neural representations.
\end{abstract}

%
%


%
%


\begin{teaserfigure}
 \includegraphics[width=\textwidth]{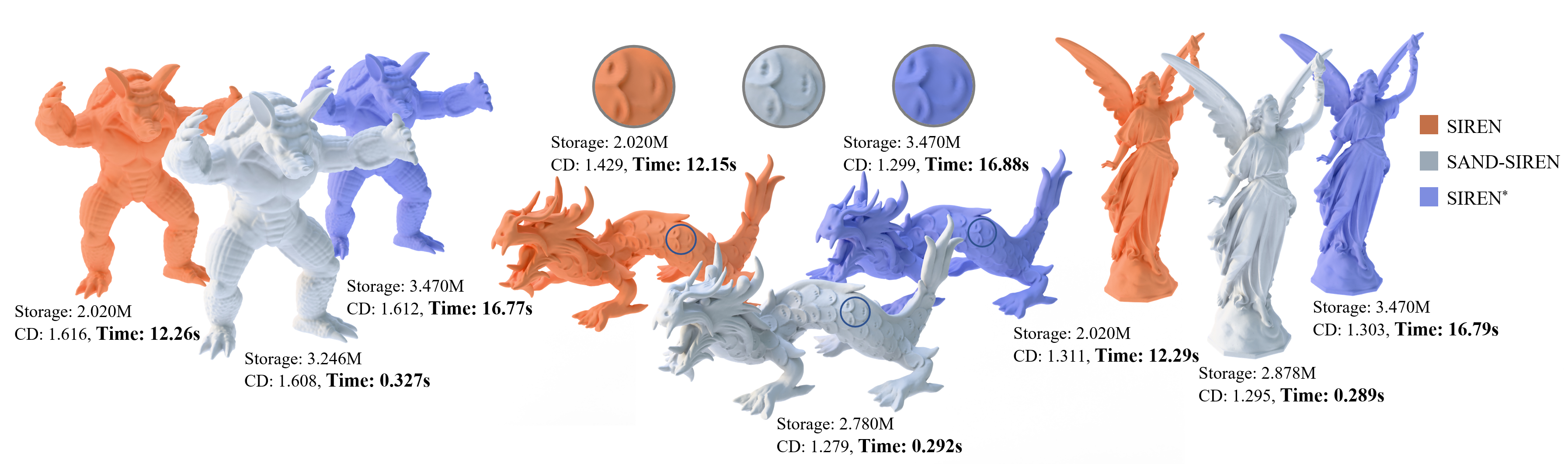}
 \caption{Unlike standard implicit neural representations (INRs) that evaluate all queries with a full network depth, SAND performs spatially adaptive evaluation. Each query is evaluated only up to the required depth determined by its spatial importance and representation complexity.
 This spatially adaptive strategy preserves geometry quality while substantially accelerating inference. Here, SIREN denotes an 8-layer network with 256 hidden units, SIREN* is a larger 8-layer 336-unit variant, and SAND-SIREN applies SAND to the 8-layer 256-unit SIREN backbone. Storage indicates model size, CD denotes Chamfer Distance, and Time measures the inference cost of querying the network during mesh extraction with Marching Cubes at a resolution of 512. The results show that applying SAND to SIREN significantly improves inference efficiency while providing high-fidelity geometric representation.}
 \label{fig:teaser}
\end{teaserfigure}

\maketitle

\section{Introduction}
Representing geometry with neural networks has emerged as an active research direction, commonly referred to as implicit neural representation (INR) \cite{park2019deepsdf,chen2021learning,wang2023lp,ren2025shape,deng2024multi}. Unlike traditional discrete representations, such as point clouds or meshes, INRs can capture intricate surface details without explicitly storing geometric primitives, providing a continuous and compact representation. \revision{Moreover, neural networks enable straightforward computation of derivatives via automatic differentiation}, making INRs particularly attractive for applications in neural rendering, 3D reconstruction, and simulation. These advantages have motivated extensive research and continuous refinement in recent years \cite{kania2024fresh,novello2025tuning,kim2023generalizable,lee2023locality,mujkanovic2024neural,Rebain_2024_CVPR,guillard2024latent,xu2025hybrid}.

Despite their strengths, INRs remain computationally demanding in practice \cite{takikawa2021neural,muller2022instant}. Obtaining accurate values for each query point typically requires evaluating the network in its full depth, and the computational cost is identical for all points. In reality, however, the precision requirements of implicit function values vary across space: points farther from the target surface require lower accuracy, while points near the surface demand higher fidelity. Even within the same iso-surface, different points may require different computational effort due to local geometric complexity.

This mismatch gives rise to a bottleneck effect: to ensure accurate representation of the zero level set and other geometrically challenging regions, networks are often globally deepened. While this satisfies the most demanding points, it simultaneously increases computation for all other regions where a shallower network would suffice. As a result, uniform network depth forces resources to be dictated by the “weakest link,” leading to substantial inefficiency and highlighting the need for spatially adaptive allocation of network computation, a feature largely absent in prior works. 

Motivated by these insights, we introduce SAND, a neural implicit geometry representation framework that employs spatially adaptive network depth. Specifically, SAND models implicit representations using a volumetric network-depth map and a tailed multi-layer perceptron (T-MLP). The volumetric network-depth map records the network evaluation depth needed in each spatial region to achieve sufficient accuracy of the implicit function. The T-MLP is a tailored multi-layer perceptron specifically designed for learning implicit functions, where an output branch, referred to as a tail, is attached to each hidden layer, enabling valid predictions to be produced at any intermediate depth. During inference, each query point first retrieves its required evaluation depth from the volumetric depth map, and the network is then executed only up to the corresponding depth. In this way, different points are evaluated with different effective network depths, allowing network computation to be adaptively allocated according to spatial importance and representational complexity, reducing redundant evaluations while preserving high-fidelity reconstruction.

This design echoes a well-established principle in traditional discrete geometric representations, where representational resources are adaptively allocated according to local geometric complexity. For instance, meshes use denser triangulations, point clouds allocate more samples, and voxel grids apply finer subdivisions in regions of high geometric complexity, while simpler regions are represented more coarsely. In a similar spirit, SAND allocates computational resources within a continuous neural implicit representation by executing deeper network evaluations in geometrically important and complex regions, while handling simpler regions with shallower evaluations.

Beyond efficiency, SAND naturally enables level-of-detail (LOD) control. By adjusting the network’s maximum evaluation depth, a single implicit representation can be queried at varying levels of geometric fidelity without requiring multiple representations or explicit decimation. This allows efficient coarse evaluations in applications where high precision is unnecessary, while progressively refining detail where higher accuracy is required. We conduct extensive experiments to validate the effectiveness of SAND and demonstrate its advantages over existing implicit neural representations.

In summary, our work makes the following contributions:
\begin{enumerate}
    \item We introduce SAND, a novel \textbf{highly-efficient} neural implicit representation framework that adaptively adjusts network evaluation depth based on spatial importance and local geometric complexity.
    
    \item We design a tailed multi-layer perceptron (T-MLP) with intermediate output branches, or tails, enabling early termination at any depth and facilitating efficient, adaptive evaluation.
    
    \item SAND naturally supports neural level of detail (LOD), allowing the same implicit representation to be queried at varying levels of geometric fidelity without multiple representations or explicit decimation.

\end{enumerate}
\section{Related Work}
Our work is closely related to research on implicit neural representations, level of detail, and efficient neural network inference. In this section, we review relevant advances in these areas.

\paragraph{Implicit Neural Representations.} Representing shapes as continuous functions using multi-layer perceptrons (MLPs) has attracted significant attention in recent years. Seminal methods encode shapes into latent codes, which are then concatenated with query coordinates and fed into a shared MLP to predict signed distances \citep{park2019deepsdf,chabra2020deep,wang2023lp}, occupancy values \citep{mescheder2019occupancy,peng2020convolutional,jiang2020local}, or unsigned distances \citep{chibane2020neural,ren2023geoudf,hu2025light}. Another line of work \citep{atzmon2020sal,gropp2020implicit,ma2020neural,ben2022digs,NEURIPS2023_2d6336c1,zhou2024cappami,yang2024monge} focuses on overfitting a single 3D shape with carefully designed regularization terms to improve surface quality. Most of these methods adopt ReLU-based MLPs, which are known to suffer from a spectral bias toward low-frequency signals. To overcome this limitation, Fourier Features \citep{tancik2020fourier} introduce a frequency-based encoding of inputs, while SIREN \citep{sitzmann2020implicit} employs periodic activation functions and specialized initialization to better capture high-frequency details. MFN \citep{fathony2021multiplicative} introduces a type of neural representation that replaces traditional layered depth with a multiplicative operation. WIRE \citep{saragadam2023wire} uses the complex Gabor wavelet as an activation function.
FINER \citep{liu2024finer} introduces an implicit neural representation with flexible spectral-bias tuning for signal representation and optimization.
Other approaches explore combining explicit feature grids such as octrees \citep{takikawa2021neural,yu2021plenoctrees} and hash tables \citep{muller2022instant} with MLPs to accelerate inference. However, these hybrid methods often incur significant memory overhead for high-fidelity geometry reconstruction. 

\begin{figure*}[th]
\centering
\begin{overpic}[width=\textwidth]{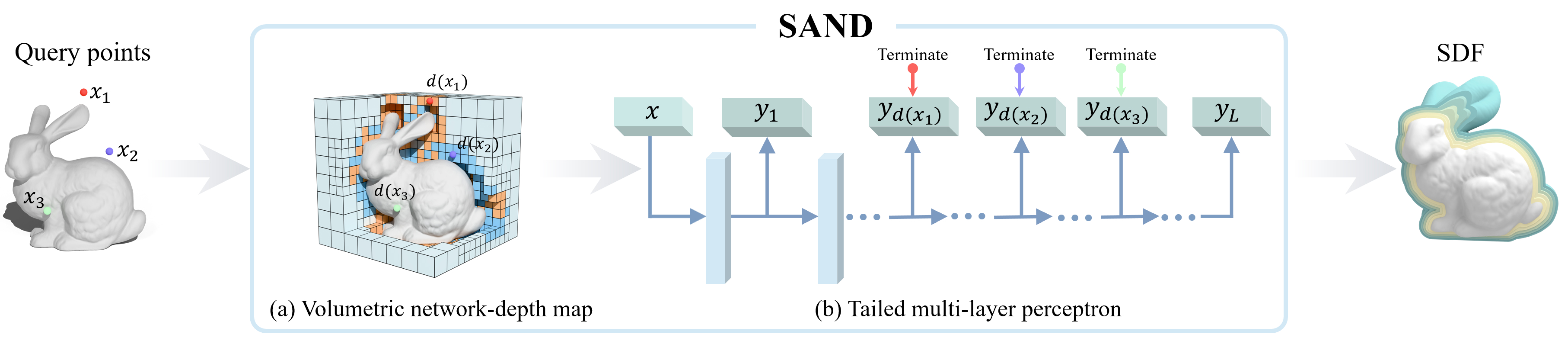}
    
\end{overpic}
\caption{Overview of SAND.
Given a query point $x$, SAND first performs a lookup in a volumetric network-depth map to determine the required evaluation depth $d(x)$ at that spatial location.
The query is then processed by a tailed multi-layer perceptron (T-MLP), which produces intermediate predictions at multiple depths.
The network evaluation terminates at depth $d(x)$, and the accumulated output up to that depth yields the predicted signed distance. By adapting the evaluation depth per query, SAND avoids redundant computation in simple and empty regions without affecting geometric accuracy.}
\label{fig:sand}
\end{figure*}

\paragraph{Level of Detail.} Level of detail (LOD) \citep{10.5555/2821571} in computer
graphics is widely used to reduce the complexity of 3D assets, aiming to improve efficiency in rendering or data transmission. Traditional geometry simplification methods \citep{10.1145/237170.237216,garland1997surface,szymczak2002piecewise,surazhsky2003explicit} focus on reducing polygon count by greedily removing mesh elements, while preserving the original mesh’s geometric characteristics to the greatest extent possible. With the rise of INRs, several methods have explored LoD modeling in implicit representations. NGLOD \citep{takikawa2021neural} and MFLOD \citep{dou2023multiplicative} leverage multilevel feature volumes to capture multiple LoDs, while PINs \citep{landgraf2022pins} introduce a progressive positional encoding scheme. BACON \citep{lindell2022bacon} proposes band-limited coordinate-based networks to represent signals at multiple scales, but its performance is sensitive to the maximum bandwidth hyperparameter. ResidualMFN \citep{shekarforoush2022residual} introduces skip connections into MFN and proposes a novel initialization method for multi-scale signal representation. \cite{mujkanovic2024neural} presents Neural Gaussian Scale-Space Fields to learn continuous, anisotropic Gaussian scale spaces directly from raw data. \cite{Rebain_2024_CVPR} proposes a novel formulation that unifies training and filtering as a maximum likelihood estimation problem, enabling neural fields to produce filtered versions of the training signal.
BANF \citep{shabanov2024banf} adopts a cascaded training strategy to train multiple \textit{independent} networks that progressively learn the residuals between the accumulated output and the ground-truth signal. However, these methods often also struggle with the high computational cost of deep networks.

\revision{ \paragraph{Efficient Neural Network Inference.} 
A large body of work has explored improving neural network efficiency through various strategies, such as pruning, quantization, and knowledge distillation, as well as dynamic inference mechanisms \citep{cheng2024survey,han2015deep,hinton2015distilling,hoefler2021sparsity,jacob2018quantization}.
Among these, early-exit methods aim to reduce inference cost via adaptive-depth evaluation \citep{rahmath2024early}.
BranchyNet \cite{teerapittayanon2016branchynet}, as a pioneering work, introduces side-branch classifiers at intermediate layers, allowing easy samples to exit early.
MSDNet \citep{huang2017multi} further improves this paradigm by adopting a multi-scale architecture with dense connectivity. 
EPNet \citep{dai2020epnet} explores a more adaptive strategy by learning exit policies through a Markov decision process, providing finer control over the inference process. E$^2$CM \citep{gormez20222} proposes a class-mean-based early-exit strategy that leverages class-wise feature statistics. \citep{jazbec2024fast} introduces statistical risk control frameworks to early-exit networks.
Despite these advances, directly applying early-exit techniques to implicit neural representations remains challenging. Implicit neural representations are inherently sensitive to approximation errors, where even minor deviations can significantly degrade representation quality. Moreover, for continuous implicit functions, it is difficult to define reliable and tractable criteria for early termination. In geometric settings, this challenge is further exacerbated by the fact that representation difficulty varies not only with geometric complexity but also with the spatial importance of different regions. 
}


\section{Proposed Method}
\label{sec-method}

\textit{Overview.}
As shown in Fig. \ref{fig:sand}, SAND consists of two main components: a volumetric network-depth map and a tailed multi-layer perceptron (T-MLP). The volumetric depth map records the network depth required for each spatial region to achieve sufficient accuracy when evaluating the implicit function. The T-MLP is a modified MLP with intermediate output branches, or tails, enabling valid predictions at any depth. During training, we first train the T-MLP to learn the implicit function. Once the T-MLP is trained, we compute the required network depth for each point according to the outputs of T-MLP. These depth values are then stored in the volumetric network-depth map for subsequent inference.
During inference, each query point first retrieves its target evaluation depth from the volumetric map, and the T-MLP is then executed only up to the corresponding depth to output implicit function values.

In the following sections, we first describe the design of the T-MLP in Section~\ref{sec-tmlp}, followed by the construction of the volumetric depth map in Section~\ref{sec-vndm}. Section~\ref{sec-training} introduces the training strategy of SAND. Finally, we present the detailed inference procedure in Section~\ref{sec-inference}.

\subsection{Tailed Multi-Layer Perceptron}
\label{sec-tmlp}
\begin{figure}[tbp]
\centering
\includegraphics[width=\linewidth]{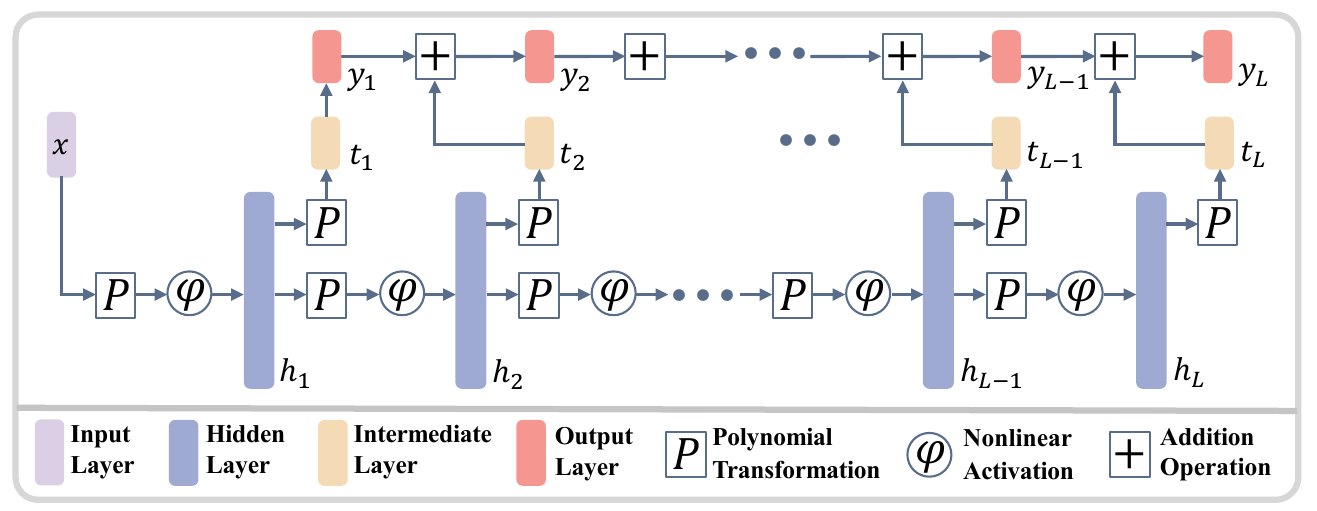}
\caption{Overview of the T-MLP architecture. \revision{Here, the Polynomial Transformation corresponds to Eq.~(\ref{multiplicative_design}).} Built on a standard MLP, the T-MLP attaches an output branch, also called a {\em tail}, after each hidden layer. The first tail produces a coarse approximation of the target function. The second tail learns the residual between the target and the first tail’s output. The third tail captures the residual between the target and the cumulative output of the first two tails. In general, the $k$-th tail models the residual between the target function and the sum of the outputs from the first $k-1$ tails.}
\label{T-MLP}
\end{figure}

The standard multi-layer perceptron (MLP) typically takes the following form:
\begin{equation}
\begin{aligned}
{h}_0 & = {x},~
{h}_{i} =\sigma\left({W}_{i} {h}_{i-1}+{b}_{i}\right), i=1, \ldots, L,~
{y}  ={W}^{out} {h}_{L}+{b}^{out},
\end{aligned}
\end{equation}
where ${x}$ is the input, $L$ denotes the number of hidden layers, ${W}_{i} \in \mathbb{R}^{N_{i} \times M_{i}}$ and ${b}_{i} \in \mathbb{R}^{N_{i}}$ define the affine transformation at the $i$-th hidden layer, and $\sigma$ denotes a nonlinear activation function. ${W}^{out}$ and ${b}^{out}$ represent the affine transformation in the output layer.

To enable early termination at any network depth, we propose the tailed multi-layer perceptron (T-MLP), as illustrated in Fig. \ref{T-MLP}. Unlike standard MLPs that produce a single output at the final layer, the T-MLP attaches an output branch, or tail, to each hidden layer. The tail at the first hidden layer is trained to produce a coarse approximation of the target function, while each subsequent tail learns the residual between the accumulated output from previous layers and the ground truth. This design allows reliable predictions at intermediate depths, facilitating adaptive evaluation according to spatial importance and complexity.  

Formally, the architecture of the T-MLP is defined as:
\begin{equation}
\label{equ:t-mlp}
\begin{aligned}
{h}_0 & = {x},~{h}_{i}  =\sigma\left({W}_{i} {h}_{i-1}+{b}_{i}\right),\\
{t}_{i} & ={W}_{i}^{out} {h}_{i}+{b}_{i}^{out},  \\
{y}_{0} & = {0},~{y}_{i} ={y}_{i-1} + {t}_{i}, i=1, \ldots, L.
\end{aligned}
\end{equation}
Here, ${t}_i$ denotes the intermediate output, i.e. residual prediction, at the $i$-th layer, and ${y}_i$ represents the accumulated output up to that layer. Each output ${y}_i$ is recursively obtained by adding the current intermediate prediction ${t}_i$ to the previous output ${y}_{i-1}$. This cumulative design enables each ${t}_i$ for $i > 1$ to focus on learning the high-frequency components not yet captured, thereby preventing redundant learning of information already accounted for by previous outputs.

Because the magnitude of the residual is typically smaller than 1, the network would struggle to train properly with such significantly small magnitudes \citep{wang2024multi}. Based on the simple fact that a value of a small magnitude can be expressed as the product of two values of larger magnitudes, we adopt a multiplicative formulation for ${t}_i$ when $i > 1$ to mitigate this issue. Specifically, we set
\begin{equation}
\begin{aligned}
{t}_{i_0} = {W}_{i_0}^{out} {h}_{i}+{b}_{i_0}^{out},~
{t}_{i_1} = {W}_{i_1}^{out} {h}_{i}+{b}_{i_1}^{out},~ 
{t}_{i} = {t}_{i_0} \circ {t}_{i_1} , i=2, \ldots, L,
\label{multiplicative_design}
\end{aligned}
\end{equation}
where $\circ$ stands for the Hadamard product, i.e., component-wise product. This multiplicative design can be interpreted as a low-rank quadratic transformation of the hidden representation ${h}_i$ to produce the output ${t}_i$, thereby enhancing the expressiveness of each output tail and improving the network’s ability to fit residuals that are challenging for purely linear output layers. A detailed proof is provided in Appendix \ref{sec:multiplicative_design}.

We denote the original loss used to train a standard MLP as $\mathcal{L}$. For our proposed T-MLP, the training objective is defined as
\begin{equation}
    \mathcal{L}_{total} = \sum_{i = 1}^{L}\mathcal{L}({y}_{i}),
\end{equation}
where ${y}_i$ denotes the cumulative output up to the $i$-th output tail.

\subsection{Volumetric Network-Depth Map}
\label{sec-vndm}
The accuracy requirement of an SDF is spatially \textit{non-uniform}: high precision is critical only near the zero level set, whereas for regions far from the surface, preserving the correct sign is often sufficient. As a result, once a T-MLP is properly trained, query points far from the surface do not need to be evaluated through the full network depth, since deeper layers contribute little to the final prediction. Furthermore, even for points on or near the zero level set, the required evaluation depth varies with local geometric complexity. In geometrically simple regions, accurate implicit values can often be obtained at intermediate layers, and the outputs of subsequent tails become negligibly small, indicating diminishing refinement. 

To model the spatially varying network depth required across different regions, we first discretize the space using an octree. The octree subdivision strategy is defined as follows: a node is further subdivided if the spatial cell corresponding to the node intersects the target shape represented by the implicit neural representation and its depth does not exceed a preset maximum depth; otherwise, the subdivision process terminates. Based on this subdivision, leaf nodes whose spatial cells intersect the target surface are labeled as near-surface regions, while all remaining leaf nodes are treated as far-from-surface regions.

Next, we define the required network depth $d(x)$ for each point as follows:
\begin{equation}
\label{equ:d(x)}
d(x)=
\begin{cases}
\min \left\{ i \;\middle|\;
S_i(x)\right\}, 
& \delta(\mathcal{O}(x)) = 0, \\[6pt]
\min \left\{ i \;\middle|\; 
\left| y_L(x) - y_i(x) \right| < r,\; S_i(x) \right\},
& \delta(\mathcal{O}(x)) = 1 .
\end{cases}
\end{equation}
where $S_i(x)$ indicates that the sign of the intermediate output $y_i(x)$ matches that of the final output $y_L(x)$, i.e.,
\[
S_i(x) := \operatorname{sign}(y_i(x)) = \operatorname{sign}(y_L(x)).
\]
$\mathcal{O}(x)$ denotes the octree leaf node containing $x$, and $\delta(\mathcal{O}(x))$ is a binary indicator, where $\delta = 0$ denotes regions far from the target surface and $\delta = 1$ denotes near-surface regions.
$y_i(x)$ is defined consistently with Eq.~(\ref{equ:t-mlp}) and represents the accumulated output of the $i$-th layer of the T-MLP.
$r>0$ is a predefined error threshold used to determine whether $y_i(x)$ is sufficiently close to the final network output.
For all points belonging to the same octree leaf node, we aggregate their required network depths by max pooling and store the maximum value at the leaf node. Formally, the network depth associated with an octree leaf ${N}$ is defined as
\begin{equation}
\label{d_leaf0}
    d({N}) = \max_{x \in {N}} d(x).
\end{equation}
\noindent

We further observe that, since points far from the shape only require the correct sign of SDF, their required network depth can be considered zero. In other words, no network evaluation is necessary, and an approximate value that preserves the correct sign can be directly stored. Based on this observation, for regions far away from the shape, we directly store this approximate value within the octree, specifically the signed distance value at the octree node center. Accordingly, we update the information stored at each octree node:
\begin{equation}
\label{d_leaf}
d({N}) =
\begin{cases}
\operatorname{sdf}(c_{N}), & \text{if } \delta({N}) = 0, \\[1em]
\displaystyle \max_{x \in {N}} d(x), & \text{if } \delta({N}) = 1,
\end{cases}
\end{equation}
where $c_{N}$ is the node center. This octree forms our final volumetric network-depth map.

\begin{figure}[tbp]
\centering
\includegraphics[width=\linewidth]{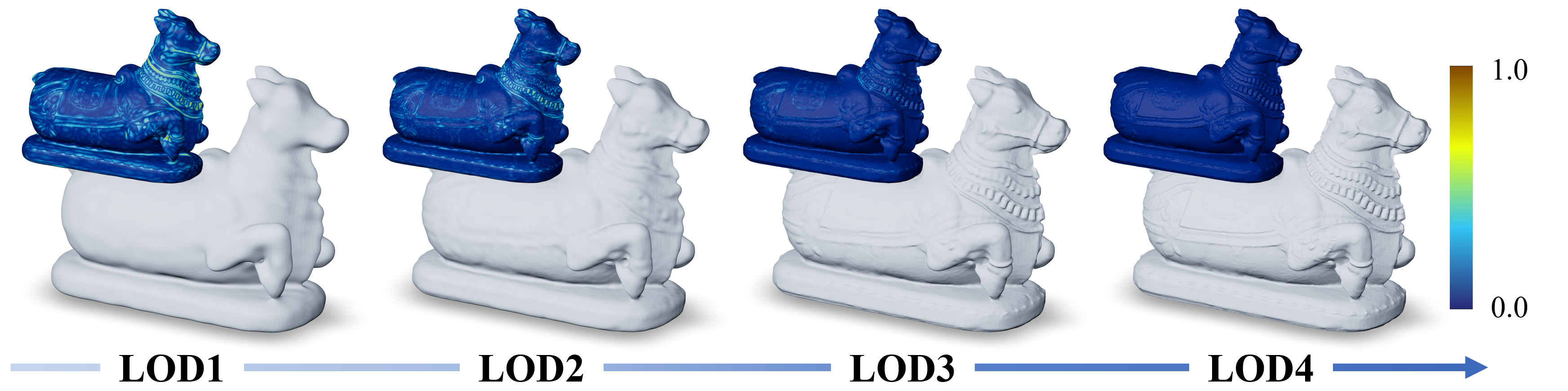}
\caption{Level of Detail. By varying the maximum network evaluation depth (1, 2, 3, 4), a single learned implicit representation produces surfaces of progressively increasing detail, corresponding to LOD1, LOD2, LOD3, and LOD4.}
\label{LOD_example}
\end{figure}

\begin{table*}[htp]
  \caption{Quantitative comparisons for 3D shape representation.
  Total time includes both octree query and network inference. For methods reported with both starred and non-starred versions, the non-starred version uses an 8-layer MLP with 256 hidden units per layer, while the starred version uses an 8-layer MLP with 336 hidden units per layer.}
  \label{table:shape_representation}
  \centering
  \resizebox{\linewidth}{!}{
  \begin{tabular}{l|c|ccc|ccc|c|ccc|ccc}
    \toprule
    \multirow{2}{*}{Method}
    & \multicolumn{7}{c|}{Stanford}
    & \multicolumn{7}{c}{Thingi10K} \\
    \cmidrule(lr){2-8} \cmidrule(lr){9-15}
    & Storage
    & \multicolumn{3}{c|}{Accuracy}
    & \multicolumn{3}{c|}{Time (s)}
    & Storage
    & \multicolumn{3}{c}{Accuracy}
    & \multicolumn{3}{c}{Time (s)} \\
    \cmidrule(lr){3-5} \cmidrule(lr){6-8}
    \cmidrule(lr){10-12} \cmidrule(lr){13-15}
    & (M)
    & CD $\downarrow$
    & F-Score $\uparrow$
    & NC $\uparrow$
    & Octree
    & Network
    & Total
    & (M)
    & CD $\downarrow$
    & F-Score $\uparrow$
    & NC $\uparrow$
    & Octree
    & Network
    & Total \\
    \midrule

    FF \citep{tancik2020fourier}
    & 2.270 & 2.432 & 92.13 & 89.84 & 0.000 & 10.81 & 10.81\,(28.6$\times$)
    & 2.270 & 3.063 & 85.92 & 93.45 & 0.000 & 10.83 & 10.83\,(26.9$\times$) \\

    FF$^*$ \citep{tancik2020fourier}
    & 3.899 & 1.651 & 94.69 & 95.51 & 0.000 & 15.02 & 15.02\,(39.7$\times$)
    & 3.899 & 1.834 & 89.75 & 97.02 & 0.000 & 15.18 & 15.18\,(37.7$\times$) \\

    SIREN \citep{sitzmann2020implicit}
    & 2.020 & 1.523 & 95.16 & 97.86 & 0.000 & 12.14 & 12.14\,(32.1$\times$)
    & 2.020 & 1.683 & 90.44 & 99.34 & 0.000 & 12.24 & 12.24\,(30.4$\times$) \\

    SIREN$^*$ \citep{sitzmann2020implicit}
    & 3.470 & 1.513 & 95.23 & 97.98 & 0.000 & 16.81 & 16.81\,(44.5$\times$)
    & 3.470 & 1.680 & 90.48 & 99.37 & 0.000 & 16.76 & 16.76\,(41.6$\times$) \\

    NGLOD \citep{takikawa2021neural}
    & 38.71 & 1.603 & 94.62 & 97.75 & 0.000 & 5.541 & 5.541\,(14.7$\times$)
    & 38.71 & 1.765 & 89.61 & 99.34 & 0.000 & 5.552 & 5.552\,(13.8$\times$) \\

    BACON \citep{lindell2022bacon}
    & 2.069 & 1.621 & 93.94 & 96.95 & 0.000 & 9.535 & 9.535\,(25.2$\times$)
    & 2.069 & 1.718 & 90.22 & 99.26 & 0.000 & 9.531 & 9.531\,(23.7$\times$) \\

    BACON$^*$ \citep{lindell2022bacon}
    & 3.530 & 1.618 & 93.96 & 96.98 & 0.000 & 16.96 & 16.96\,(44.9$\times$)
    & 3.530 & 1.709 & 90.25 & 99.29 & 0.000 & 16.88 & 16.88\,(41.9$\times$) \\

    BANF \citep{shabanov2024banf}
    & 7.905 & 1.870 & 89.09 & 94.82 & 0.000 & 45.26 & 45.26\,(120$\times$)
    & 7.905 & 4.417 & 74.13 & 96.19 & 0.000 & 45.30 & 45.30\,(112$\times$) \\

    FINER \citep{liu2024finer}
    & 2.020 & 1.520 & 95.19 & 97.91 & 0.000 & 21.62 & 21.62\,(57.2$\times$)
    & 2.020 & 1.683 & 90.44 & 99.35 & 0.000 & 21.60 & 21.60\,(53.6$\times$) \\

    FINER$^*$ \citep{liu2024finer}
    & 3.470 & 1.518 & 95.24 & 97.99 & 0.000 & 29.14 & 29.14\,(77.1$\times$)
    & 3.470 & 1.680 & 90.47 & 99.38 & 0.000 & 29.21 & 29.21\,(72.5$\times$) \\
    
    \midrule

    SAND-SIREN$^-$
    & 3.095 & 1.512 & 95.23 & 98.26 & 0.096 & 0.296 & 0.392\,(1.04$\times$)
    & 3.500 & \textbf{1.671} & 90.51 & 99.49 & 0.097 & 0.354 & 0.451\,(1.12$\times$) \\

    SAND-FINER$^-$
    & 3.120 & \textbf{1.498} & \textbf{95.30} & \textbf{98.29} & 0.101 & 0.402 & 0.503\,(1.33$\times$)
    & 3.511 & \textbf{1.671} & \textbf{90.52} & \textbf{99.50} & 0.100 & 0.513 & 0.613\,(1.52$\times$) \\

    \midrule

    SAND-SIREN
    & 3.095 & 1.512 & 95.23 & 98.26 & 0.096 & 0.214 & \textbf{0.310}\,(0.82$\times$)
    & 3.500 & \textbf{1.671} & 90.51 & 99.49 & 0.097 & 0.247 & \textbf{0.344}\,(0.85$\times$) \\

    SAND-FINER
    & 3.120 & \textbf{1.498} & \textbf{95.30} & \textbf{98.29} & 0.101 & 0.277 & {0.378}\,(1.00$\times$)
    & 3.511 & \textbf{1.671} & \textbf{90.52} & \textbf{99.50} & 0.100 & 0.302 & {0.403}\,(1.00$\times$) \\

    \bottomrule
  \end{tabular}}
\end{table*}

\revision{ \textbf{Remark.} In the octree construction,  ``the target shape represented by the implicit neural representation'' does not necessarily refer to the ground-truth shape associated with the implicit neural representation. For tasks where ground-truth geometry is available (e.g., shape overfitting), the octree can be constructed directly using the ground-truth shape.
For tasks without ground-truth geometry, we first extract a mesh from the trained implicit neural representation via zero-level set extraction, and construct the octree based on this mesh.
This process does not introduce additional error, as the extracted mesh is consistent with the shape encoded by the implicit neural representation itself. Moreover, it does not incur additional inference cost, since it is performed as a post-processing step.}

\subsection{Training Procedure}
\label{sec-training}

\revision{For tasks where ground-truth geometry is available (e.g., shape overfitting), we first construct an octree using the ground-truth shape prior to training. For regions far from the shape, the required network depth is considered to be zero, and only the approximate value that preserves the correct sign needs to be stored. Therefore, the leaf nodes corresponding to regions far from the shape can be determined before T-MLP training and directly assigned SDF values at their centers. During training, these distant regions are ignored and the T-MLP is then trained on regions near the surface. After training, the leaf nodes near the shape are updated according to Eqs. (\ref{equ:d(x)}) and (\ref{d_leaf}), storing the maximum effective network depth among points in each node.}

\revision{For tasks without ground-truth geometry (e.g., surface reconstruction from point clouds), we first train the T-MLP following the standard pipeline. After training, we extract a mesh from the trained T-MLP via zero-level set extraction, and then construct the octree based on this mesh as described in Section~\ref{sec-vndm}.}
\subsection{Inference Procedure}
\label{sec-inference}
During inference, we first query the volumetric network-depth map (octree) for all points. If a query falls into a leaf node marked as being far from the zero level set, the required network depth is zero and no network evaluation is performed; the SDF value is approximated using the value stored at the corresponding octree node. For the remaining points, we perform network evaluation with point-wise adaptive depths according to the queried network depth, yielding accurate SDF values.

Moreover, due to the multi-level outputs of the T-MLP, intermediate layers may not fully capture fine geometric details and thus naturally provide coarser representations. As shown in Fig. \ref{LOD_example}, by restricting the maximum network evaluation depth during inference, we can explicitly control the level of detail (LOD) of the reconstructed geometry. As a result, our method naturally supports LOD representation.

\section{Experiments}

\label{sec:experiment}
In this section, we evaluate the effectiveness of SAND on shape overfitting tasks. We first provide the implementation details and clarify the evaluation metrics used in our experiments. We then present comparative results between SAND and baseline methods on 3D shape representation and neural level of detail (LOD) tasks to demonstrate its effectiveness. Finally, we conduct a series of ablation studies using the Stanford 3D Scanning Repository to validate the contribution of each design component and to examine the effects of hyperparameters.

\begin{figure*}[htbp]
\centering
\begin{overpic}[width=\textwidth]{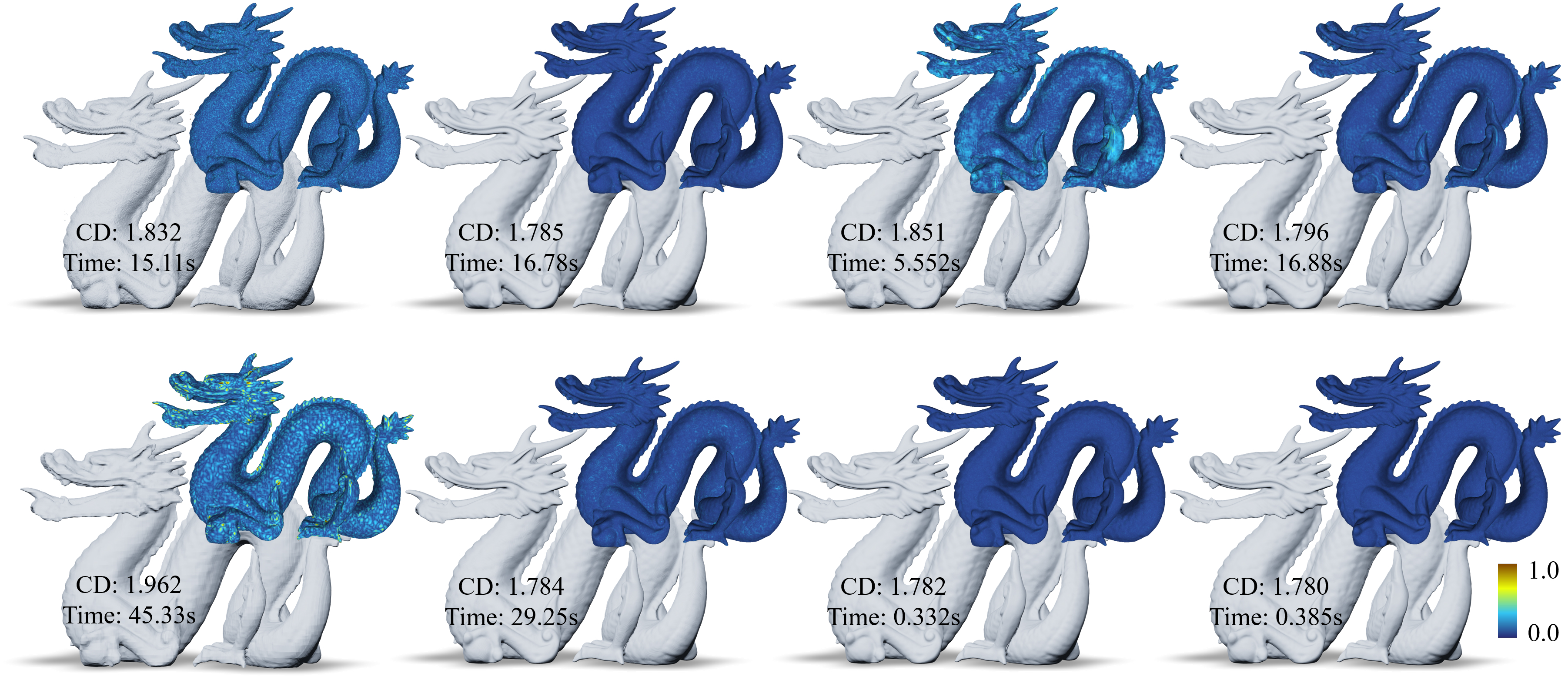}
    \put(11.0,21.2){FF$^*$}
    \put(34.5,21.2){SIREN$^*$}
    \put(57.7,21.2){NGLOD}
    \put(82.2,21.20){BACON$^*$}
    
    \put(10.0, -1.2){BANF}
    \put(34.5, -1.2){FINER$^*$}
    \put(55.7, -1.2){SAND-SIREN}
    \put(80.4, -1.27){SAND-FINER}
\end{overpic}
\caption{Visual comparisons for 3D shape representation.}
\label{fig:3d_representation}
\end{figure*}
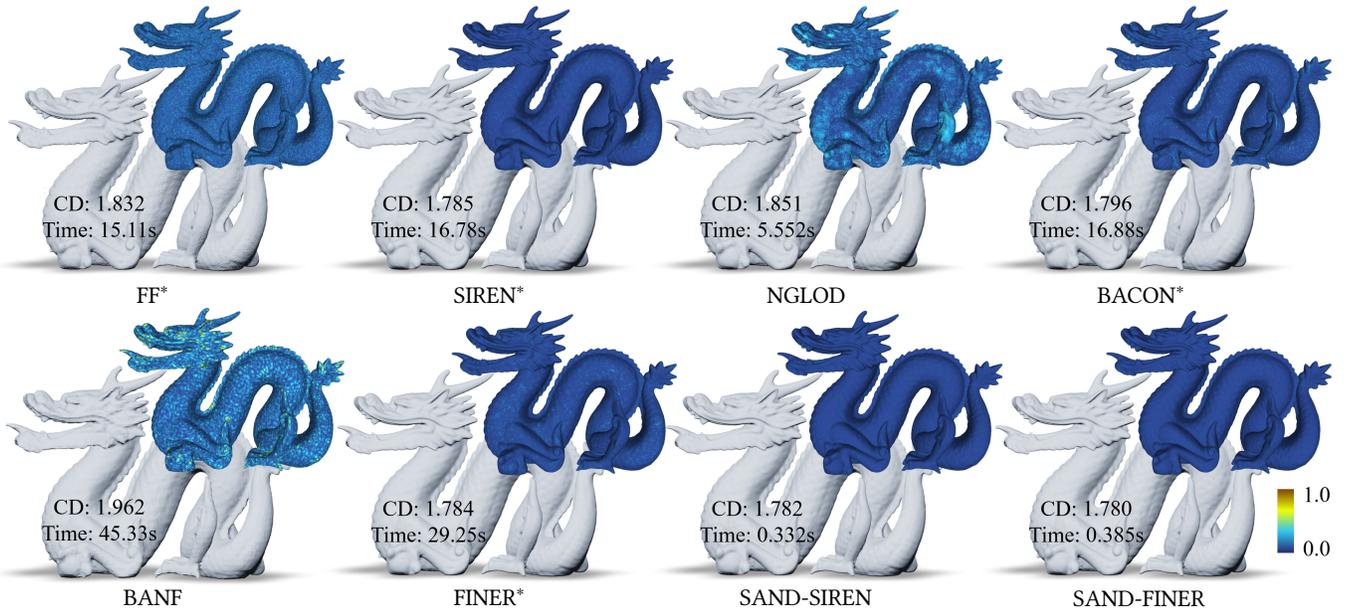

\subsection{Experimental Setup}
\paragraph{Implementation Details} We implement SAND in Python and conduct all experiments on an NVIDIA RTX 4090 GPU with 24GB video memory and an Intel Xeon Platinum 8457C CPU. We use 3D models from the Thingi32 subset of Thingi10K \citep{zhou2016thingi10k} and the Stanford 3D Scanning Repository. T-MLP, configured with 8 hidden layers of 256 units each, is employed to fit the SDF. The maximum depth of volumetric network-depth map (octree) is set to 9. 
\revision{We determine whether a voxel of the octree intersects the mesh by checking if the distance from its center to the mesh is less than or equal to half of the voxel’s diagonal length.}
The loss is formulated as:
\begin{equation}
    \mathcal{L}_{sdf} = \sum_{i=1}^{L} \frac{1}{\left| \mathcal{Q} \right|} \sum_{{x} \in \mathcal{Q}} \left| y_{i}({x}) - y_{gt}({x}) \right|,
\end{equation}
where ${y}_{i}$ denotes the cumulative output up to the $i$-th output tail of T-MLP, ${y}_{gt}$ denotes the ground-truth SDF value, and $\mathcal{Q}$ represents the set of sampled query points. 
\revision{We use the Open3D library to compute the ground-truth SDF.}
The Adam optimizer is used with the learning rate of $1 \times 10^{-4}$ and training is run for 100k iterations. We sample 100k training points. Among them, 60\% are surface points, and the remaining 40\% are obtained by perturbing the surface points with Gaussian noise ($\sigma = 0.01$). The octree is queried to filter out points belonging to regions far from the zero level set, so that network training focuses on regions near the target surface.
The error threshold $r$ is set to 0.00015. \revision{All shapes are normalized to a unit cube so that the error threshold is applicable across objects of different scales.} We extract meshes from the SDFs using the Marching Cubes (MC) algorithm \citep{lorensen1998marching} with a grid resolution of $512^3$. 

\paragraph{Evaluation Metrics.} To evaluate representation accuracy, we uniformly sample 500k points from each extracted mesh and compute the Chamfer Distance (CD), Normal Consistency (NC), and F-Score. The reported CD uses L1-norm and is scaled by $10^3$. The NC, expressed as a percentage,
refers to the mean absolute cosine of normals in one
mesh and normals at nearest neighbors in the other mesh.
The F-Score, expressed as a percentage, indicates the harmonic
mean of precision and recall, with a default threshold
of 0.003.
To evaluate inference efficiency, we report the query time including both octree query and network evaluation when calculating the $512^3$
grid during the MC process.

\begin{figure}[tbp]
\centering
\includegraphics[width=\linewidth]{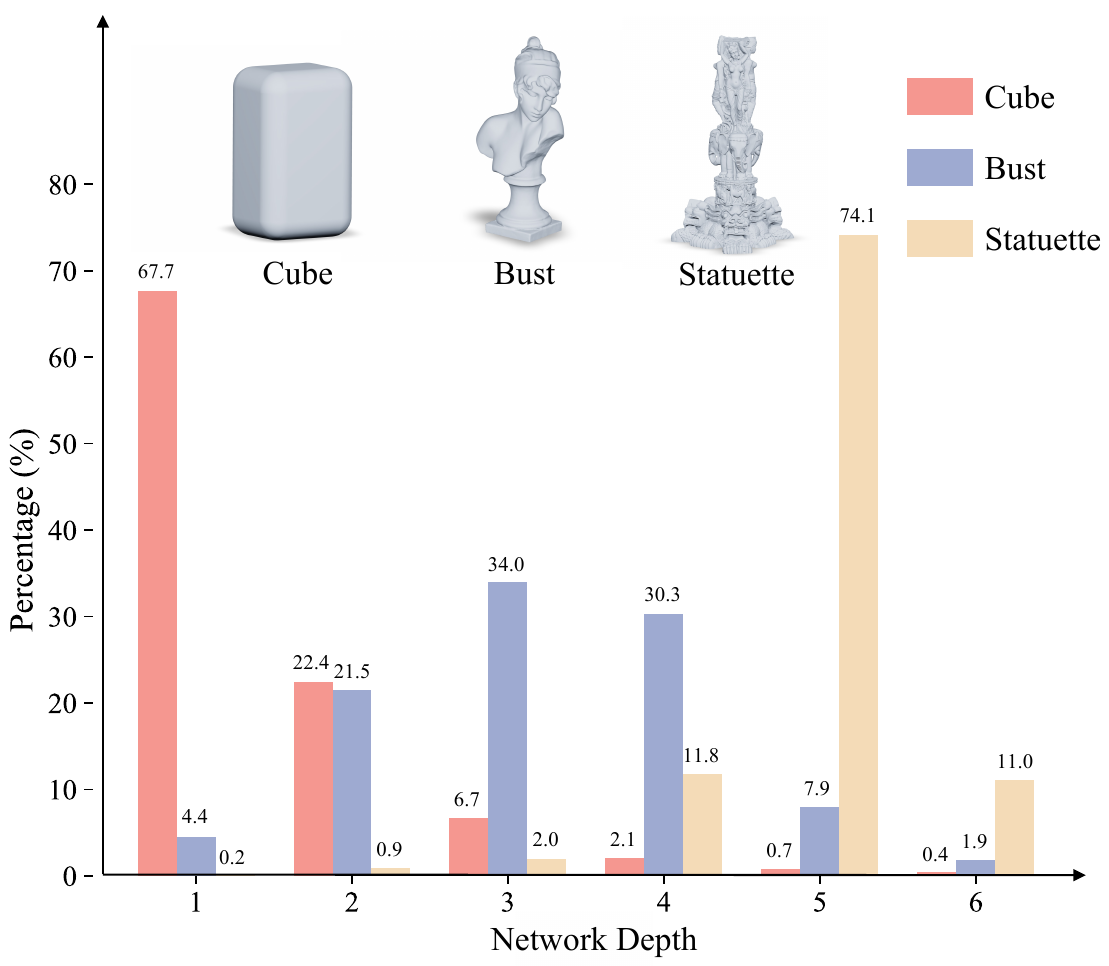}
\caption{Distribution of network depth usage for shapes with different geometric complexity under our SAND framework.}
\label{depth_distribution}
\end{figure}

\subsection{Efficient 3D Shape Representation}
To evaluate the effectiveness of SAND, we conduct experiments on two representative MLP architectures, SIREN \citep{sitzmann2020implicit} and FINER \cite{liu2024finer}, resulting in SAND-SIREN and SAND-FINER, respectively. We also evaluate ablated variants that only apply adaptive depth assignment in the far field, where points far from the zero level set are assigned
a depth of zero, while disabling adaptive depth near the surface. These variants are denoted as SAND-SIREN$^{-}$ and SAND-FINER$^{-}$.

We compare our method with several state-of-the-art baselines, including Fourier Features (FF) \citep{tancik2020fourier}, SIREN \citep{sitzmann2020implicit}, NGLOD \citep{takikawa2021neural}, BACON \citep{lindell2022bacon}, BANF \citep{shabanov2024banf}, and FINER \citep{liu2024finer}.  
For FF, SIREN, BACON, and FINER, we evaluate two network configurations: an 8-layer MLP with 256 hidden units per layer, which matches the network architecture used in our method, and a larger 8-layer MLP with 336 hidden units per layer, whose storage footprint is comparable to that of our network combined with the octree. We denote the latter configuration with a superscript $^{*}$, while results without the superscript correspond to the former setting. Since BANF has not released its official implementation for the 3D shape representation task, we reimplement BANF for this task based on the description provided in the paper.

We report quantitative and qualitative comparisons in Tab. \ref{table:shape_representation} and Fig. \ref{fig:3d_representation}. As shown in these results, both SAND-SIREN and SAND-FINER significantly improve the inference efficiency compared to their corresponding baselines, SIREN and FINER, and also outperform other competing methods. The efficiency gain arises from SAND’s spatially adaptive network evaluation strategy, which applies different network depths to different spatial regions, rather than evaluating all points with the full network depth, thereby saving computation in less important and geometrically simple areas.
Meanwhile, SAND also surpasses the baseline methods in representation accuracy, as it allows the network to concentrate its representational capacity on regions near the target shape, which are the most critical for accurate surface representation.
However, since SAND requires an additional octree to store network-depth information, it inevitably introduces extra storage overhead. Nevertheless, the reconstruction accuracy of our method still surpasses that of the baselines when their storage is increased to a comparable level.

In addition, SAND-SIREN and SAND-FINER produce the same representation results as their corresponding variants, SAND-SIREN$^{-}$ and SAND-FINER$^{-}$. This indicates that applying adaptive network depths in regions close to the target shape improves inference efficiency without compromising representation accuracy.

\begin{table}[tbp]
  \caption{\revision{Statistics of absolute residuals of T-MLP tails at different depths for shapes with varying geometric complexity.}}
  \label{table:residual_statistics}
  \centering
  \resizebox{\linewidth}{!}{
  \begin{tabular}{lcccc|cccc}
    \toprule
    \multirow{2}{*}{\revision{Shape}}  & \multirow{2}{*}{\revision{Depth}} & \revision{Max.} & \revision{Mean} & \revision{Med.}
          & \multirow{2}{*}{\revision{Depth}} & \revision{Max.} & \revision{Mean} & \revision{Med.} \\
          &   & \revision{($\times10^{-3}$)} & \revision{($\times10^{-4}$)} & \revision{($\times10^{-4}$)}
          &   & \revision{($\times10^{-3}$)} & \revision{($\times10^{-4}$)} & \revision{($\times10^{-4}$)} \\
    \midrule

    \revision{Cube}      & \revision{1} & \revision{16.78} & \revision{69.23} & \revision{69.02} & \revision{2} & \revision{5.360} & \revision{0.815} & \revision{0.384} \\
    \revision{Bust}      & \revision{1} & \revision{21.88} & \revision{51.10} & \revision{43.52} & \revision{2} & \revision{15.98} & \revision{15.58} & \revision{10.32} \\
    \revision{Statuette} & \revision{1} & \revision{24.81} & \revision{32.04} & \revision{20.32} & \revision{2} & \revision{20.28} & \revision{31.46} & \revision{21.36} \\
    \midrule

    \revision{Cube}      & \revision{3} & \revision{2.944} & \revision{0.354} & \revision{0.225} & \revision{4} & \revision{0.326} & \revision{0.220} & \revision{0.134} \\
    \revision{Bust}      & \revision{3} & \revision{15.62} & \revision{5.426} & \revision{2.646} & \revision{4} & \revision{10.65} & \revision{1.617} & \revision{1.232} \\
    \revision{Statuette} & \revision{3} & \revision{16.43} & \revision{11.02} & \revision{5.252} & \revision{4} & \revision{11.68} & \revision{3.735} & \revision{2.397} \\
    \midrule

    \revision{Cube}      & \revision{5} & \revision{0.192} & \revision{0.114} & \revision{0.068} & \revision{6} & \revision{0.157} & \revision{0.051} & \revision{0.031} \\
    \revision{Bust}      & \revision{5} & \revision{1.629} & \revision{0.472} & \revision{0.266} & \revision{6} & \revision{0.183} & \revision{0.088} & \revision{0.049} \\
    \revision{Statuette} & \revision{5} & \revision{4.259} & \revision{2.296} & \revision{1.708} & \revision{6} & \revision{1.368} & \revision{0.836} & \revision{0.576} \\
    \bottomrule
  \end{tabular}}
\end{table}

\subsection{Analysis of Network Depth Distribution}
To better understand how SAND allocates computation across different spatial regions, we analyze the distribution of network depth used during inference. Fig.~\ref{depth_distribution} presents histograms showing, under our SAND framework, the proportion of points evaluated at each layer of a six-layer network for three representative shapes: rounded cube, bust, and statuette, corresponding to low, medium, and high geometric complexity, respectively. For the simple rounded cube, a large fraction of points are processed by the lower layers, with only a small number of points requiring deeper network evaluations, which enables SAND to achieve substantial acceleration. In contrast, as geometric complexity increases, a higher proportion of points are propagated to deeper layers, reflecting the greater representational demand in regions with fine geometric details. Consequently, the achievable acceleration of SAND decreases for more complex shapes.

\revision{We also report the statistics of absolute residuals of T-MLP tails at different depths for shapes with varying geometric complexity in Tab.~\ref{table:residual_statistics}.
It can be observed that the first tail (depth = 1) produces relatively large outputs, as it is responsible for approximating the SDF itself. In contrast, subsequent tails yield progressively smaller outputs, as they learn the residuals between the current prediction and the ground truth, which decrease as the approximation becomes more accurate with increasing depth.
For simple shapes (e.g., a cube), the outputs of deeper tails become negligible from early depths, indicating that many tails are redundant. In contrast, for more complex shapes, fewer tails exhibit negligible outputs, suggesting that more tails are required to refine the underlying geometry. This observation is consistent with the intuition that higher geometric complexity demands greater model capacity.}

In conclusion, these results demonstrate that SAND automatically allocates computational effort according to local geometric complexity, concentrating deeper evaluations near complex regions while avoiding unnecessary computation in simpler areas.

\subsection{Comparison with Instant-NGP}
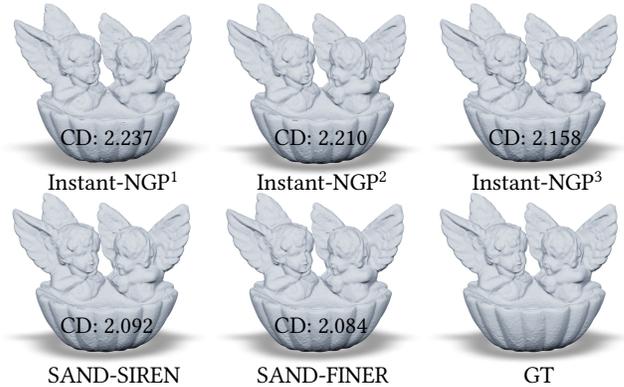
\begin{figure}[tbp]
\centering
\begin{overpic}[width=\linewidth]{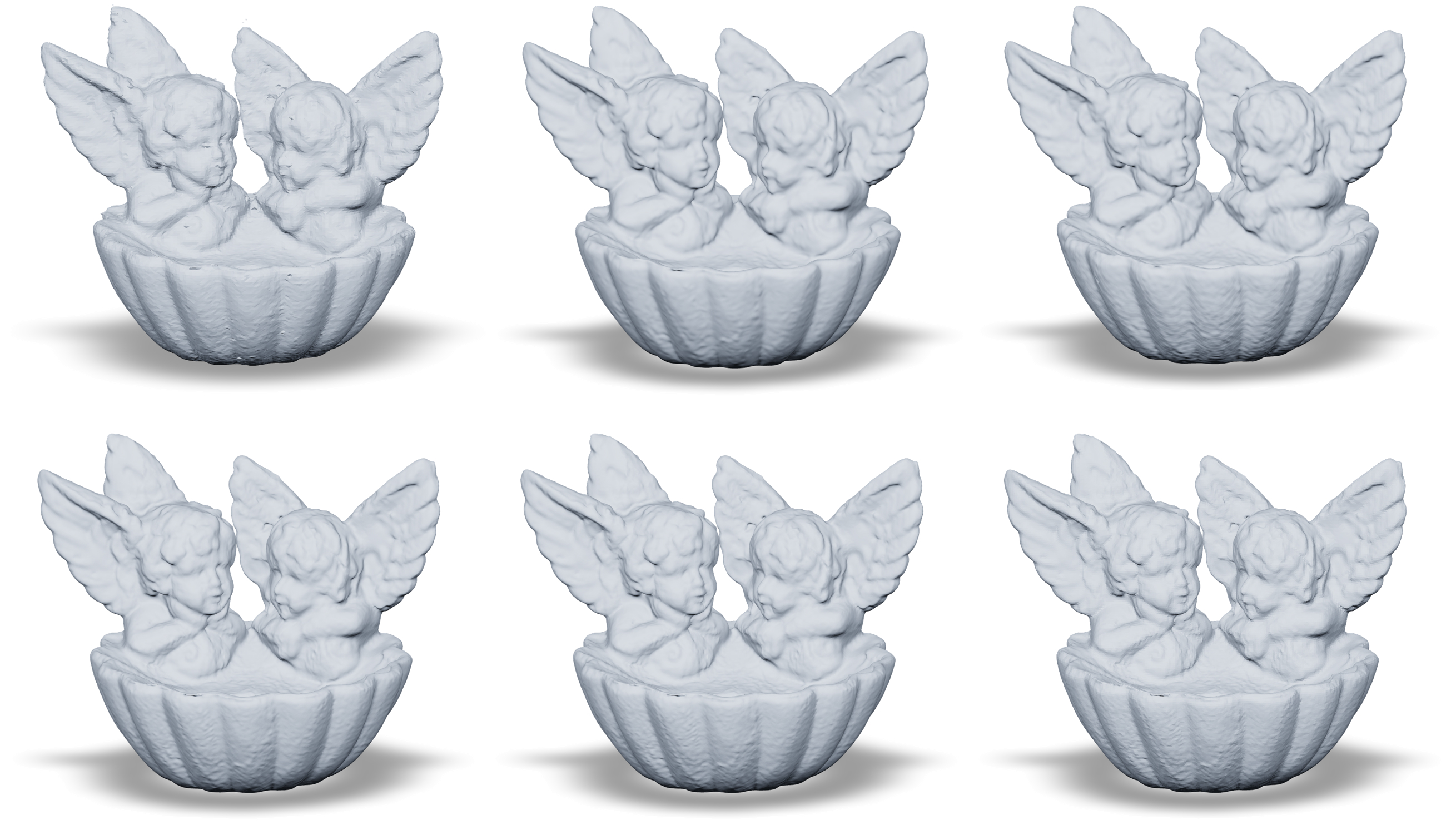}

    \put(9.9, 35.0){CD: 2.237}
    \put(43.2, 35.0){CD: 2.210}
    \put(76.5, 35.0){CD: 2.158}
    
    \put(9.9, 5.8){CD: 2.092}
    \put(43.2, 5.8){CD: 2.084}

    \put(8.0, 28.0){Instant-NGP$^1$}
    \put(40.5, 28.0){Instant-NGP$^2$}
    \put(74.0, 28.0){Instant-NGP$^3$}
    
    \put(8.0, -2.0){SAND-SIREN}
    \put(40.5, -2.0){SAND-FINER}
    \put(82.0, -2.0){GT}
    
\end{overpic}
\caption{\revision{Visual comparisons with Instant-NGP.}}
\label{fig:InstantNGP}
\end{figure}

Instant-NGP \citep{muller2022instant} is a highly efficient framework for neural implicit representations that leverages a multiresolution hash grid to store learned features and combines it with a small MLP. Its implementation is highly optimized, written directly in CUDA, and heavily leverages GPU parallelism. \revision{We compare our method with Instant-NGP models of different sizes on the Thingi10k dataset, and report the results in Fig.~\ref{fig:InstantNGP} and Tab.~\ref{table:instant-NGP}.} As shown, our method achieves significantly higher representation accuracy.

Despite providing substantial acceleration over conventional INRs, SAND still exhibits a speed gap compared to Instant-NGP. This gap is, to some extent, attributed to implementation differences rather than inherent algorithmic complexity. Instant-NGP is highly optimized in CUDA and specifically tailored for GPU execution, whereas our method is implemented in PyTorch, which naturally leads to slower inference.
\revision{The FLOPs reported in Tab.~\ref{table:instant-NGP} further indicate that the two methods have comparable computational complexity.} Instant-NGP achieves its efficiency through a space–time trade-off by storing large multiresolution feature grids in memory, thereby significantly reducing computation at inference time. While this design enables extremely fast queries, it may limit representational accuracy under constrained memory budgets.

In contrast, SAND adopts an adaptive computation strategy that allocates network capacity according to spatial importance and local geometric complexity. This allows SAND to achieve a favorable balance between efficiency and accuracy without relying on heavy memory consumption. As a result, SAND maintains high reconstruction fidelity even under limited storage, and its design is flexible and can be readily integrated with different MLP architectures, such as SIREN or FINER.

\begin{table}[tbp]
  \centering
  \caption{\revision{Quantitative comparisons with Instant-NGP.}}
  \label{table:instant-NGP}
  \resizebox{\linewidth}{!}{
  \begin{tabular}{c|cccccc}
    \toprule
    Method
      & Storage (MB)
      & CD $\downarrow$ 
      & F-score $\uparrow$ 
      & NC $\uparrow$ 
      & Time (s)
      & FLOPs \\
    \midrule
    Instant-NGP$^1$
      & 3.823
      & 1.823
      & 88.55
      & 98.23
      & \textbf{0.012}
      & \textbf{$\mathbf{1.68\times10^{12}}$}\\
    \revision{Instant-NGP$^2$}
      & \revision{5.187}
      & \revision{1.801}
      & \revision{89.02}
      & \revision{98.77}
      & \revision{\textbf{0.012}}
      & \revision{$1.73\times10^{12}$}\\
    \revision{Instant-NGP$^3$}
      & \revision{7.302}
      & \revision{1.736}
      & \revision{89.78}
      & \revision{99.12}
      & \revision{{0.013}}
      & \revision{$1.84\times10^{12}$} \\
    SAND-SIREN
      & 3.500
      & \textbf{1.671}
      & 90.51
      & 99.49
      & 0.344
      & $1.74\times10^{12}$\\
    SAND-FINER
      & 3.511
      & \textbf{1.671}
      & \textbf{90.52}
      & \textbf{99.50}
      & 0.403
      & $1.76\times10^{12}$\\
    \bottomrule
  \end{tabular}}
\end{table}

\begin{table}[tp]
  \caption{Quantitative comparisons for 3D shape representation across multiple levels of detail (LODs).}
  \label{table:lod_representation}
  \centering
  \resizebox{\linewidth}{!}{
  \begin{tabular}{c|l|ccc|ccc}
    \toprule
    \multirow{2}{*}{LoD} 
    & \multirow{2}{*}{Method} 
    & \multicolumn{3}{c|}{Stanford} 
    & \multicolumn{3}{c}{Thingi10K} \\
    \cmidrule{3-8}
     & 
     & CD $\downarrow$ 
     & F-Score $\uparrow$
     & NC $\uparrow$ 
     & CD $\downarrow$ 
     & F-Score $\uparrow$ 
     & NC $\uparrow$ \\
    \midrule

    \multirow{5}{*}{LOD4}
      & NGLOD       & 1.603 & 94.62 & 97.75 & 1.765 & 89.61 & 99.34 \\
      & BACON$^*$   & 1.618 & 93.96 & 96.98 & 1.709 & 90.25 & 99.29 \\
      & BANF        & 1.870 & 89.09 & 94.82 & 4.417 & 74.13 & 96.19\\
      & SAND-SIREN  & 1.512 & 95.23 & \textbf{98.29} & 1.671 & 90.51 & 99.49 \\
      & SAND-FINER  & \textbf{1.498} & \textbf{95.30} & \textbf{98.29} & \textbf{1.671} & \textbf{90.52} & \textbf{99.50} \\
    \midrule

    \multirow{5}{*}{LOD3}
      & NGLOD       & 1.729 & 92.73 & 96.82 & 1.799 & 89.25 & 99.15 \\
      & BACON$^*$   & 1.904 & 88.65 & 95.43 & 1.795 & 89.11 & 98.99 \\
      & BANF        & 1.906 & 87.80 & 94.24 & 4.172 & 74.52 & 96.28 \\
      & SAND-SIREN  & 1.549 & 94.76 & 97.48 & 1.681 & 90.46 & 99.41 \\
      & SAND-FINER  & \textbf{1.519} & \textbf{95.17} & \textbf{97.89} & \textbf{1.675} & \textbf{90.50} & \textbf{99.46} \\
    \midrule

    \multirow{5}{*}{LOD2}
      & NGLOD       & 2.044 & 85.25 & 94.96 & 1.975 & 85.65 & 98.65 \\
      & BACON$^*$   & 2.430 & 77.64 & 93.30 & 1.966 & 84.00 & 97.81 \\
      & BANF        & 2.785 & 69.19 & 89.72 & 6.208 & 65.87 & 93.82 \\
      & SAND-SIREN  & 1.778 & 90.36 & 95.56 & 1.739 & 89.58 & 99.06 \\
      & SAND-FINER  & \textbf{1.580} & \textbf{94.24} & \textbf{97.14} & \textbf{1.687} & \textbf{90.38} & \textbf{99.37} \\
    \midrule

    \multirow{5}{*}{LOD1}
      & NGLOD       & 2.743 & 68.26 & 92.17 & 2.417 & 74.50 & 97.64 \\
      & BACON$^*$   & 4.253 & 54.83 & 88.65 & 2.858 & 70.27 & 96.05 \\
      & BANF        & 5.061 & 44.39 & 83.19 & 8.089 & 48.65 & 90.91 \\
      & SAND-SIREN  & 3.369 & 61.91 & 87.94 & 2.446 & 76.40 & 96.34 \\
      & SAND-FINER  & \textbf{2.085} & \textbf{84.22} & \textbf{93.63} & \textbf{1.836} & \textbf{87.72} & \textbf{98.59} \\
    \bottomrule
  \end{tabular}}
\end{table}
\begin{figure*}[htbp]
\centering
\begin{overpic}[width=\textwidth]{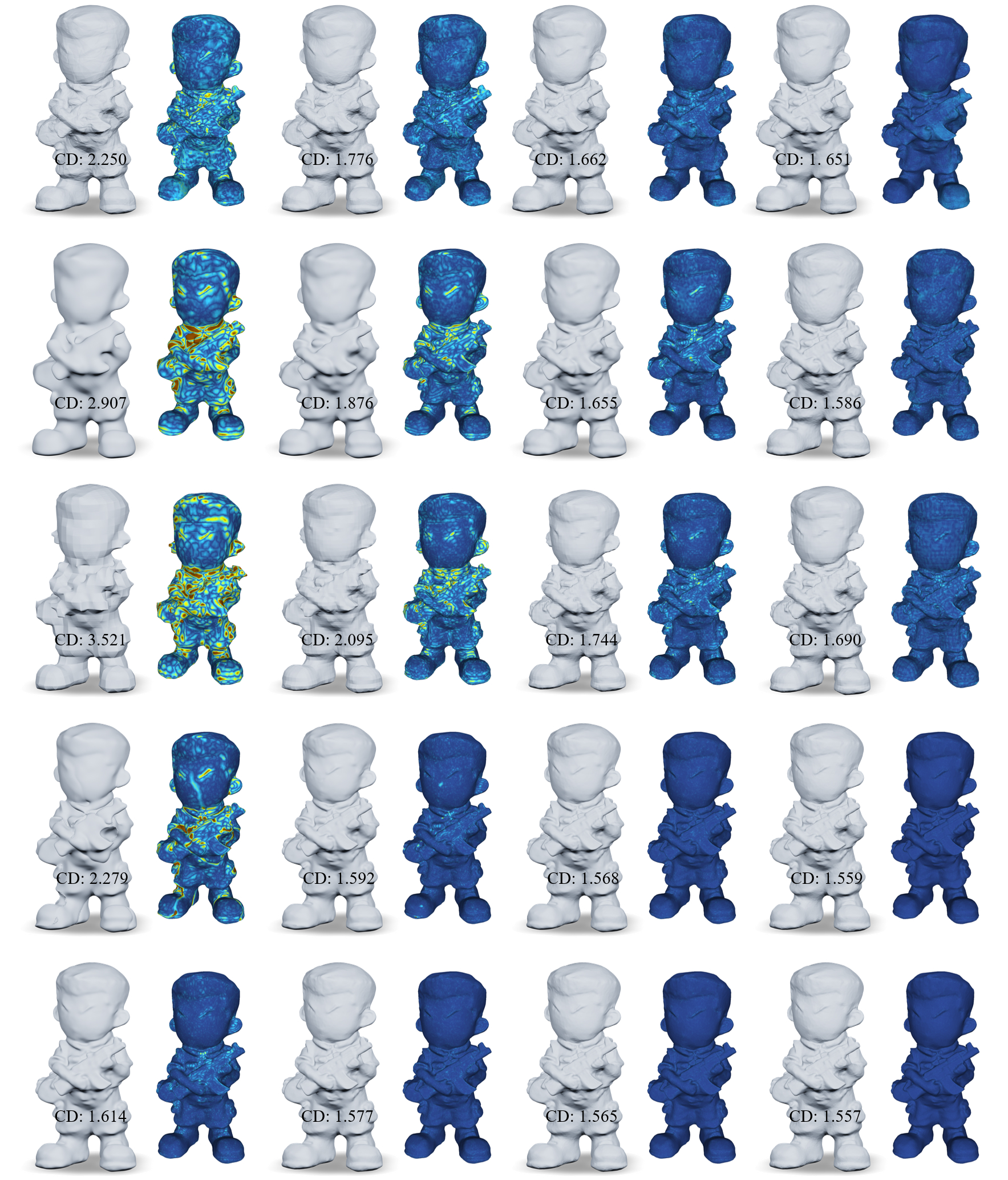}
    \put(7.6,80.0){NGLOD / LOD1}
    \put(28.5,80.0){NGLOD / LOD2 }
    \put(49.0,80.0){NGLOD / LOD3 }
    \put(69.7, 80.0){NGLOD / LOD4 }
    
    \put(7.3, 59.5){BACON$^*$ / LOD1}
    \put(28.2, 59.5){BACON$^*$ / LOD2}
    \put(48.7, 59.5){BACON$^*$ / LOD3}
    \put(69.4, 59.5){BACON$^*$ / LOD4}

    \put(8.0,39.3){BANF / LOD1}
    \put(29.0,39.3){BANF / LOD2}
    \put(49.5,39.3){BANF / LOD3}
    \put(70.3,39.3){BANF / LOD4}
    
    \put(5.5, 19.0){SAND-SIREN / LOD1}
    \put(26.50, 19.0){SAND-SIREN / LOD2}
    \put(47.50, 19.0){SAND-SIREN / LOD3}
    \put(68.50, 19.0){SAND-SIREN / LOD4}

    \put(5.5, -1.2){SAND-FINER / LOD1}
    \put(26.5, -1.2){SAND-FINER / LOD2}
    \put(47.5, -1.2){SAND-FINER / LOD3}
    \put(68.5, -1.2){SAND-FINER / LOD4}
    
\end{overpic}
\caption{Visual comparisons for 3D shape LOD representation.}
\label{fig:LOD}
\end{figure*}
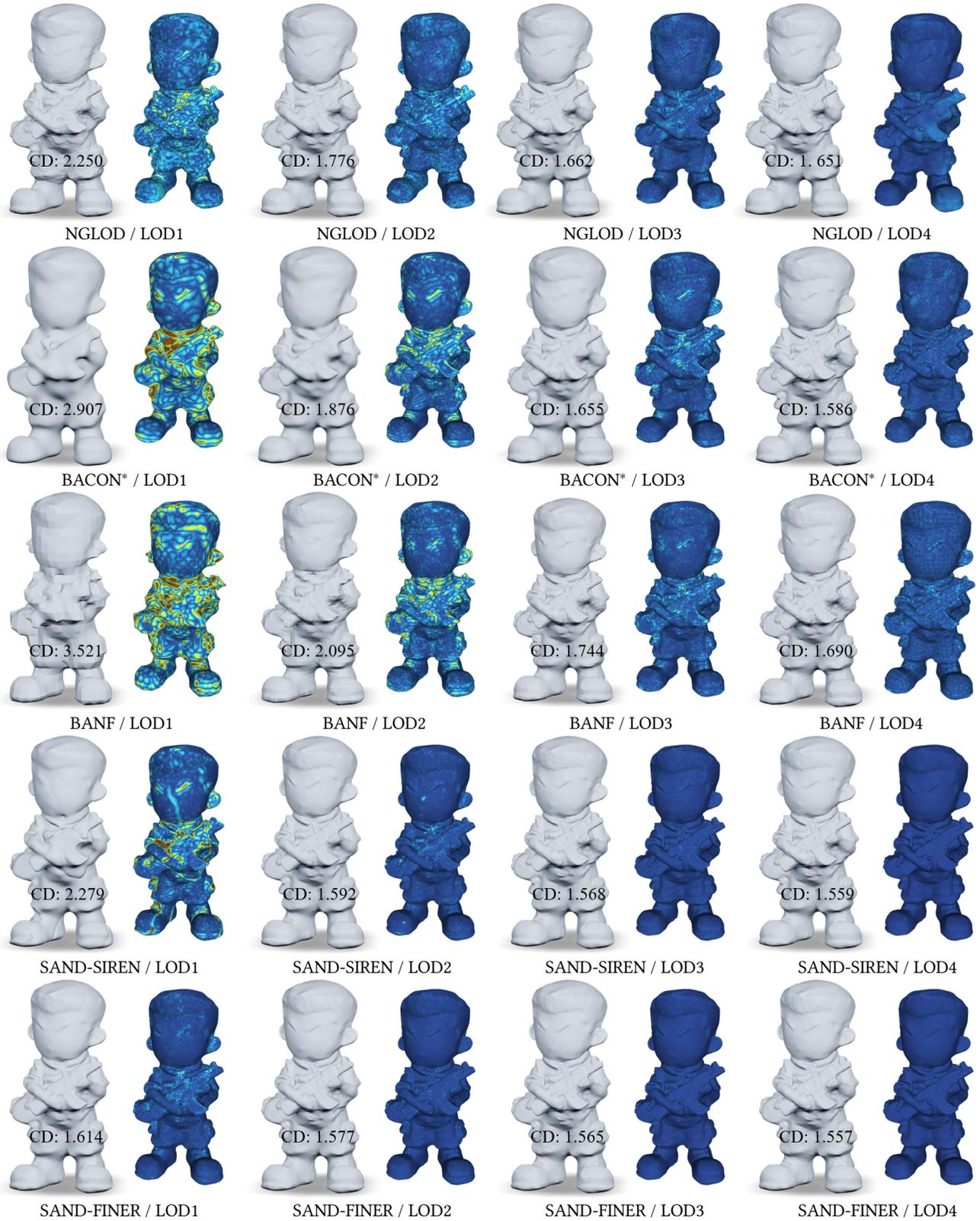

\subsection{Neural Levels of Detail}
LoD is critical for applications that require adaptive resolution, such as rendering acceleration and model compression. We also evaluate our method under the level-of-detail (LOD) task and, following BACON, report results at maximum network depths of 2, 4, 6, and 8, corresponding to LOD1, LOD2, LOD3, and LOD4, respectively. As shown in Fig.~\ref{fig:LOD} and Tab.~\ref{table:lod_representation}, both SAND-SIREN and SAND-FINER consistently outperform the baseline methods. NGLOD requires a large number of parameters to achieve satisfactory shape representation. For BACON, we observe that its performance is highly sensitive to the maximum bandwidth hyperparameter: a small value leads to overly smooth shapes, while a large value results in rough and irregular geometry. BANF incurs high computational costs due to querying multiple $N^3$ grids at different resolutions and struggles to capture shape features.

\subsection{Ablation Studies}

\begin{figure}[ht]
\centering
\includegraphics[width=\linewidth]{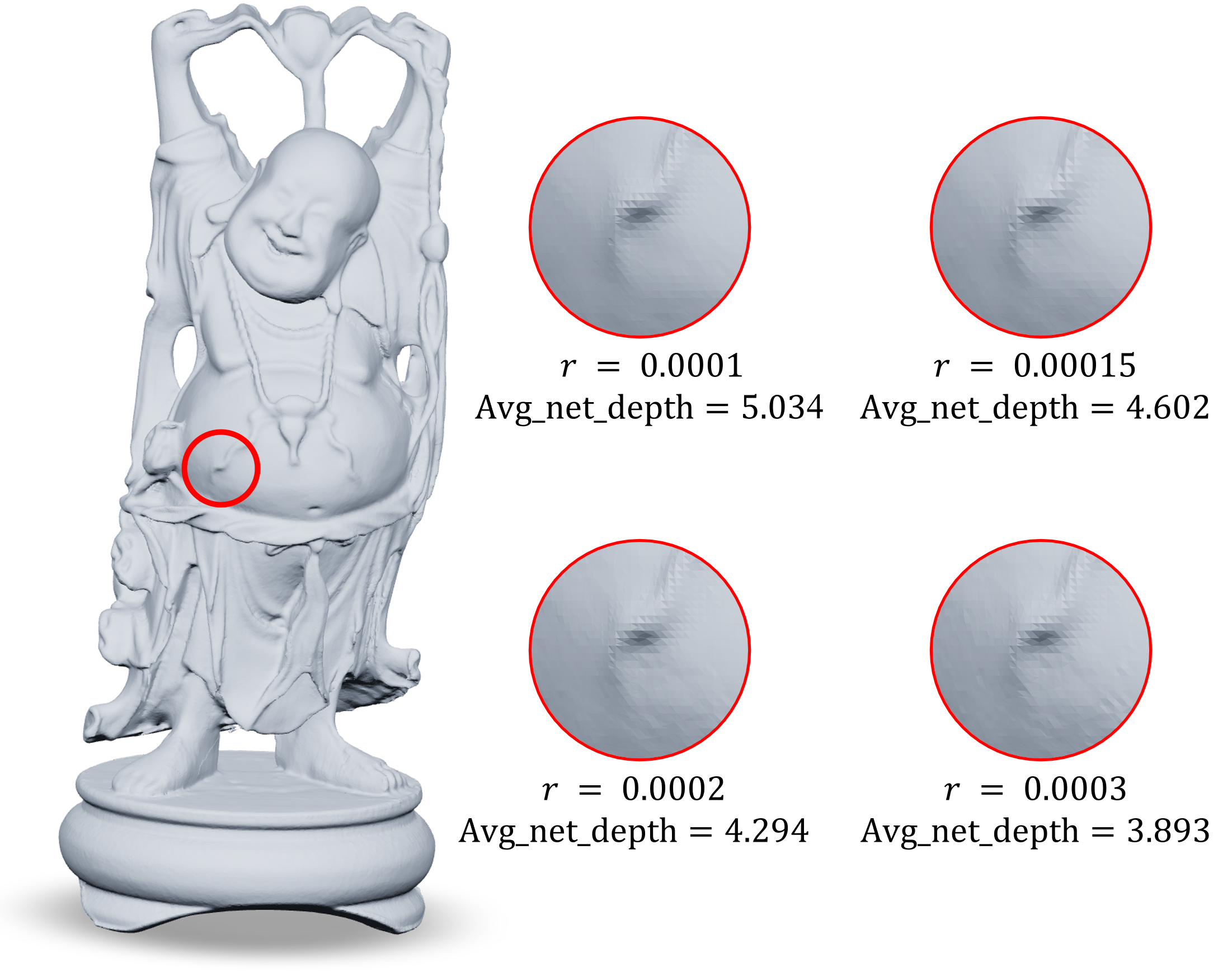}
\caption{Effect of the error threshold.}
\label{error_threshold}
\end{figure}

\paragraph{Error Threshold} We vary the error threshold $r$ in Eq.~(\ref{equ:d(x)}) to study its effect on representation accuracy and acceleration performance. As shown in Fig.~\ref{error_threshold}, we set $r$ to $0.0001$, $0.00015$, $0.0002$, and $0.0003$, respectively. As $r$ increases, the average network evaluation depth during inference decreases, leading to improved acceleration. However, when $r = 0.0002$, the extracted surfaces exhibit slight imperfections. To balance reconstruction accuracy and acceleration performance, we set $r = 0.00015$ in all experiments.

\paragraph{Octree Depth} We set the maximum octree depth to 6, 7, 8, 9, and 10 to evaluate the effect of different maximum octree depths on acceleration performance. As shown in Tab. \ref{table:octree_depth}, increasing the maximum octree depth leads to better acceleration, since finer spatial partitioning allows more accurate assignment of adaptive network depths. However, a deeper octree also results in higher storage overhead. To balance acceleration performance and storage cost, we set the maximum octree depth to 9 in all experiments. 

\begin{table}[htbp]
  \centering
  \caption{Effect of the maximum octree depth.}
  \label{table:octree_depth}
  \begin{tabular}{c|ccccc}
    \toprule
    Max Depth
      & 6
      & 7
      & 8
      & 9
      & 10 \\
    \midrule
    Time (s) 
      & 0.886
      & 0.600
      & 0.468
      & 0.378
      & 0.332 \\
    Storage (MB)
      & 2.052
      & 2.085
      & 2.230
      & 3.120
      & 7.938 \\
    \bottomrule
  \end{tabular}
\end{table}

\revision{\paragraph{Necessity of the Octree Structure} We further conduct experiments to validate the necessity and advantages of the octree structure in the SAND framework. To this end, we remove the octree and instead adopt a dynamic inference strategy, where the decision to proceed to the next layer is made based on the output at the current depth. Specifically, if the absolute value of the predicted residual at the current layer falls below the error threshold $r$, the computation is terminated; otherwise, it proceeds to the next layer until the maximum depth is reached. We report the quantitative comparisons in Tab.~\ref{table:octree}. It can be observed that, without the octree structure, both inference time and representation accuracy are inferior to those of the full SAND model with the octree.}

\revision{This performance gap arises from several factors. First, without constructing the octree prior to inference, it is not possible to identify regions far from the surface and determine their correct SDF signs. As a result, the same termination strategy must be applied to both near-surface and distant points, although such high precision is unnecessary for regions far from the surface.
Second, dynamically evaluating whether the predicted residual falls below the threshold 
$r$ during inference introduces additional computational overhead.
Third, near the zero level set, the SDF values are inherently small. Even if the residual at the current layer falls below the threshold $r$, it may still significantly affect the final prediction and can even alter the sign of the output.
Finally, without the octree, training requires sampling points across the entire domain, rather than focusing on regions near the surface where accurate modeling is most critical.
In contrast, although the octree incurs only a modest increase in memory usage, it enables highly efficient depth determination, supports zero-network-depth queries for unimportant regions, and avoids the runtime overhead associated with dynamic inference strategies.}

\begin{table}[tbp]
  \centering
  \caption{\revision{Ablation study on the octree structure.}}
  \label{table:octree}
  \resizebox{\linewidth}{!}{
  \begin{tabular}{l|c|c|c|c|c}
    \toprule
    \revision{Network} 
    & \revision{Storage (MB)}
    & \revision{CD $\downarrow$} 
    & \revision{F-score $\uparrow$} 
    & \revision{NC $\uparrow$}
    & \revision{Time (s)} \\
    \midrule
    \revision{SAND w/o Octree}   & \revision{\textbf{2.041}}   & \revision{1.521} & \revision{95.19} & \revision{97.89} & \revision{13.67} \\
    \revision{Full SAND (Ours)}  & \revision{3.120}   & \revision{\textbf{1.498}} & \revision{\textbf{95.30}} & \revision{\textbf{98.29}} & \revision{\textbf{0.378}} \\
    \bottomrule
  \end{tabular}}
\end{table}

\paragraph{Residual Design of T-MLP} To evaluate the effectiveness of the residual design in T-MLP, we make each output tail directly learn the ground truth rather than learning the residual, and conduct experiments using the Stanford 3D Scanning Repository. The quantitative comparisons in Tab.~\ref{table:multiplicative_design} show that T-MLP without the residual design is less effective than our version with it. This is because the residual formulation enables the later hidden representations to focus on learning the residuals between the current approximation and the ground-truth signal, avoiding redundantly learning the information already encoded by earlier layers.

\paragraph{Multiplicative Design of T-MLP} We conduct experiments to verify the effectiveness of the multiplicative design in Eq. (\ref{multiplicative_design}). As illustrated in Tab. \ref{table:multiplicative_design}, incorporating the multiplicative design leads to more accurate 3D shape representations compared to the baseline without it.
\revision{We further replace the multiplicative design with a learnable scaler to validate its effectiveness. The quantitative results in Tab.~\ref{table:multiplicative_design} show that, although introducing a learnable scaler improves performance, it is still inferior to the multiplicative design. This is because a learnable scaler is difficult to initialize differently across layers prior to training, which is crucial for stable and effective optimization.
In contrast, the proposed multiplicative design enhances the network’s representational capacity with minimal additional parameters and computational overhead, while being supported by a solid theoretical foundation. It provides stronger learning capability than a learnable scaler.}
\begin{table}[htbp]
  \centering
  \caption{\revision{Effect of the residual and multiplicative designs. 
  ``Mult.'' denotes multiplicative interaction and ``Scaler'' refers to a learnable scaler modulation.}}
  \label{table:multiplicative_design}
  
  \begin{tabular}{l|c|c|c}
    \toprule
    Network 
    & CD $\downarrow$ 
    & F-score $\uparrow$ 
    & NC $\uparrow$ \\
    \midrule
    T-MLP w/o Residual Design      & 1.547 & 95.02 & 97.85 \\
    T-MLP w/o Mult.          & 1.509 & 95.26 & 98.08 \\
    \revision{T-MLP w/o Mult. + Scaler} & \revision{1.505} & \revision{95.26} & \revision{98.15} \\
    Full T-MLP (Ours)               & \textbf{1.498} & \textbf{95.30} & \textbf{98.29} \\
    \bottomrule
  \end{tabular}
\end{table}

\paragraph{Sampling Strategy} To evaluate the effectiveness of training the network only on regions near the zero level set, we combine our volumetric network-depth map with a conventional MLP and train it using our sampling strategy, that is, by sampling points only near the zero level set. This framework differs from SAND in that for near-surface regions, the network depth stored in the former is set to the maximum depth for all points, whereas in SAND it is stored as an adaptive depth. Both approaches set the network depth to zero for regions far from the zero level set. We report results on the Stanford Scan dataset using FINER in Tab. \ref{table:sampling_strategy}. Here, FINER denotes the conventional framework without a volumetric network-depth map, FINER$^+$ refers to the conventional MLP combined with a volumetric network-depth map, and SAND-FINER represents our proposed framework. FINER$^+$ outperforms FINER, demonstrating the effectiveness of our sampling strategy. However, FINER$^+$ still falls short of SAND-FINER. We attribute our improvement to the residual and multiplicative design in our framework, as well as the fact that the T-MLP supervises all hidden layers, enabling more stable and effective optimization, rather than relying solely on backpropagation to indirectly adjust the parameters of earlier layers.
\begin{table}[htbp]
  \centering
  \caption{Effect of the sampling strategy.}
  \label{table:sampling_strategy}
  \begin{tabular}{l|c|c|c}
    \toprule
    Network 
    & CD $\downarrow$ 
    & F-score $\uparrow$ 
    & NC $\uparrow$ \\
    \midrule
    FINER       & 1.520 & 95.19 & 97.91 \\
    FINER$^+$   & 1.502 & 95.28 & 98.18 \\
    SAND-FINER  & \textbf{1.498} & \textbf{95.30} & \textbf{98.29} \\
    \bottomrule
  \end{tabular}
\end{table}
\paragraph{Training Time}
During training of the T-MLP, a loss is computed for each output, which inevitably increases the time required for a single training iteration. In Tab.~\ref{table:training_time}, we report the per-iteration training time for both an 8-layer 256-hidden-unit conventional MLP and the T-MLP.
\begin{table}[htbp]
  \centering
  \caption{Timing costs per training iteration.}
  \label{table:training_time}
    \resizebox{\linewidth}{!}{

  \begin{tabular}{c|ccccc}
    \toprule
      & SIREN (MLP)
      & SIREN (T-MLP)
      & FINER (MLP)
      & FINER (T-MLP) \\
    \midrule
    Time (s) 
      & 0.0129
      & 0.0241
      & 0.0283
      & 0.0416 \\
    \bottomrule
  \end{tabular}}
\end{table}

\section{Conclusion and Future Work}
\label{sec:conclusion}
In this paper, we presented SAND, a neural implicit geometry representation framework with spatially adaptive network depth that explicitly aligns computational effort with spatial importance and local geometric complexity. By combining a volumetric network-depth map with a tailed multi-layer perceptron (T-MLP), SAND enables adaptive network evaluation, allowing computation to terminate early in simple or low-importance regions while allocating deeper evaluations to geometrically complex areas near the surface. This design not only significantly improves inference efficiency, but also naturally supports level-of-detail (LOD) control from a single implicit representation, without requiring multiple networks or explicit decimation. Extensive experiments on 3D shape representation and neural LOD tasks demonstrate that SAND consistently outperforms existing implicit neural representations in both inference efficiency and reconstruction quality.

At present, SAND determines adaptive network depth after network training. A promising direction for future work is to explore whether adaptive depths can be predicted prior to or jointly with training based on local geometric complexity, which could further extend the benefits of adaptive computation to the training stage. Moreover, extending SAND beyond per-shape volumetric depth maps toward more generalizable depth prediction mechanisms remains an important avenue for future research.

In summary, by introducing spatially adaptive computational resource allocation into implicit neural representations, SAND moves toward more efficient, flexible, and resource-aware implicit modeling. We believe this perspective opens up promising directions for future research on adaptive, real-time, and scalable geometric representation learning.

\bibliographystyle{ACM-Reference-Format}
\bibliography{bibliography}

\clearpage
\appendix
\section{Appendix}

\setcounter{figure}{0}
\setcounter{table}{0}
\setcounter{equation}{0}
\subsection{Multiplicative Design of Tailed Multi-Layer Perceptron}
\label{sec:multiplicative_design}
The multiplicative design defined in Eq. (\ref{multiplicative_design}) of the main paper is given as:
\begin{equation}
\begin{aligned}
{t}_{i_0} &= {W}_{i_0}^{out} {h}_{i}+{b}_{i_0}^{out},  \\
{t}_{i_1} &= {W}_{i_1}^{out} {h}_{i}+{b}_{i_1}^{out}, \\
{t}_{i} &= {t}_{i_0} \circ {t}_{i_1} , i=2, \ldots, L,
\end{aligned}
\end{equation}
where ${W}^{out}_{i_0} \in \mathbb{R}^{D \times N_{i}}$, ${W}^{out}_{i_1} \in \mathbb{R}^{D \times N_{i}}$, ${b}^{out}_{i_0} \in \mathbb{R}^{D}$ and ${b}^{out}_{i_1} \in \mathbb{R}^{D}$. Here, $D$ is the dimension of output ${t}_{i}$ and $N_{i}$ denotes the dimension of the $i$-th hidden representation ${h}_{i}$. For clarity, consider the case where the output $t_i$ is a scalar. Let ${a}^\top = W^{out}_{i_0} \in \mathbb{R}^{1 \times N_{i}} $,  $ {b}^\top = W^{out}_{i_1} \in \mathbb{R}^{1 \times N_{i}} $,  $ {x} = {h}_i \in \mathbb{R}^{N_{i} \times 1} $, $ c={b}_{i_0}^{out} \in \mathbb{R} $ and $ d={b}_{i_1}^{out} \in \mathbb{R} $. Then the output $t_i$ can be rewritten as:
\begin{equation}
    t_i = ({a}^\top {x} + c)({b}^\top {x} + d) = ({a}^\top {x})({b}^\top {x}) + d ({a}^\top {x}) + c ({b}^\top {x}) + cd.
\end{equation}

Alternatively, this expression can be written in compact matrix form as:
\begin{equation}
    t_i = {x}^\top Q {x} + {u}^\top {x} + s,
\end{equation}
where  $Q = {a} {b}^\top \in \mathbb{R}^{N_{i} \times N_{i}} $,  ${u}^\top = d {a}^\top + c {b}^\top \in \mathbb{R}^{1 \times N_{i}} $,  and $ s = cd \in \mathbb{R} $.

This formulation shows that T-MLP implements a low-rank quadratic transformation of the hidden representation ${x} $ (i.e., $ {h}_i $) to produce the output  $t_i$. In the case where $ t_i $ is multi-dimensional, the same operation is applied independently to each output dimension.
\end{document}